\documentclass[prd,nofootinbib,]{revtex4}
\usepackage{dcolumn}
\usepackage{graphicx}
\usepackage{amssymb}
\usepackage{lscape}
\usepackage{amsmath}
\usepackage{psfrag}
\usepackage{epsfig}
\usepackage{footnote}
\usepackage{subfigure}
\usepackage{float}
\usepackage{hyperref}

\begin{document}

\def\be{\begin{equation}}
\def\ee{\end{equation}}
\def\bea{\begin{eqnarray}}
\def\eea{\end{eqnarray}}
\def\ba{\begin{array}}
\def\ea{\end{array}}
\def\bse{\begin{subequations}}
\def\ese{\end{subequations}}

\def\pd{{\partial}}

\def\a{{\alpha}}
\def\b{{\beta}}
\def\g{{\gamma}}
\def\m{{\mu}}
\def\n{{\nu}}
\def\d{{\delta}}
\def\e{{\epsilon}}
\def\vp{{\varphi}}
\def\o{{\omega}}
\def\k{{\kappa}}
\def\l{{\lambda}}

\def\K{{\mathcal{K}}}
\def\U{{\mathcal{U}}}
\def\M2{{M_P^2}}
\def\R{\mathcal{R}}
\def\L{\mathcal{L}}
\def\vpc{{\dot{\varphi}_\text{cosm}}}
\newcommand{\sch}{Schwarzschild }
\newcommand{\V}{Vainshtein }
\newcommand{\Stuc}{Stuckelberg }
{\renewcommand{\arraystretch}{2.5}\renewcommand{\tabcolsep}{0.4cm}
\long\def\symbolfootnote[#1]#2{\begingroup
\def\thefootnote{\fnsymbol{footnote}}\footnote[#1]{#2}\endgroup}

\title{Restoring General Relativity in massive bi-gravity theory}

\author{Eugeny Babichev$^{a}$, Marco Crisostomi$^{a,b,c,d}$}
\affiliation{$^a$ Univ. Paris-Sud, Laboratoire de Physique Th\'eorique, CNRS UMR 8627, F-91405 Orsay, France}
\affiliation{$^b$ Dipartimento di Scienze Fisiche e Chimiche, Universit\`a di L'Aquila, I-67010 L'Aquila, Italy}
\affiliation{$^c$ INFN, Laboratori Nazionali del Gran Sasso, I-67010 Assergi, Italy}
\affiliation{$^d$ Universit\'e Paris Diderot - Paris 7, F-75205 Paris, France}

\begin{abstract}
We study static spherically symmetric solutions of massive bi-gravity theory, free from the Boulware-Deser ghost. 
We show the recovery of General Relativity via the Vainshtein mechanism, in the weak limit of the physical metric. 
We find a single polynomial equation determining the behavior of the solution for distances smaller than the inverse graviton mass.
This equation is generically of the seventh order, while for a specific choice of the parameters of the theory it can be reduced to lower orders.
The solution is analytic in different regimes: for distances below the Vainshtein radius (where General Relativity is recovered), and in the opposite regime, beyond the Vainshtein radius, where the solution approaches the flat metric.
\end{abstract}

\date{\today}

\maketitle 

\section{Introduction}
Modification of General Relativity (GR) giving mass to the graviton started from the work of Fierz and Pauli~\cite{Fierz:1939ix}: 
they considered a linear theory of a single massive spin-2 field living in flat spacetime. 
The first nonlinear realization of the massive graviton was presented much later~\cite{Isham:1971gm}, although in a completely different context.
To extend at the non-perturbative level the action of Fierz and Pauli, adding to the Einstein-Hilbert
action a non-derivative self-coupling for the metric $g$, it is required the
introduction of an additional metric $f$ that may be a fixed external field, or a dynamical one.
When $f$ is non-dynamical we are in the framework of 
\ae ther-like theories where diff. invariance can be restored by the introduction of a suitable set of \Stuc fields; on the
other hand, if it is dynamical, we enter in the context of bigravity theories.

Unfortunately, for a generic potential the theory has a ghost propagating degree of freedom (d.o.f.)~\cite{Boulware:1973my}, 
the so-called Boulware-Deser ghost, associated with the Ostrogradski ghost in more general setup. 
Notably, the Fierz-Pauli theory was constructed so that it has five healthy propagating degrees of freedom, 
while the sixth mode is removed due to the specific choice of the coefficients in the mass term.
When the theory is promoted to nonlinear level or considered around non-flat background, the sixth mode reappears leading to ghost instability. 
This problem was solved only recently by a careful choice of the massive gravity potential~\cite{deRham:2010ik}
such that, on fully non perturbative level, the theory propagates only five degrees of freedom~\cite{Hassan:2011hr} (see also \cite{Hassan:2011vm,Comelli:2012vz,Deffayet:2012nr}).
We will refer to this theory as the de Rham-Gabadadze-Tolley (dRGT) model.
Of course, one has to be cautious about the rest --- the five propagating degrees of freedom --- however, at least there is 
no {\it a priori} Ostrogradski instability associated with the sixth mode.
The original dRGT construction of massive gravity takes
the additional metric as a flat, non-dynamical field;
then it was extended in the bi-gravity context supplying an extra Einstein-Hilbert term for the second metric~\cite{Hassan:2011zd}.
The bi-metric approach to massive gravity is the main subject of our work. 
We would like to emphasize here that the bi-gravity formulation of massive gravity is not just a theoretical  entertainment, but
also cosmology calls for it.
When the second metric is non-dynamical and Minkowski 
there is no homogeneous spatially flat FRW solution~\cite{D'Amico:2011jj, Gumrukcuoglu:2011ew}, 
on the contrary in the bigravity formulation flat FRW homogeneous solutions do
exist~\cite{Comelli:2011zm,vonStrauss:2011mq,Volkov:2011an}.
Moreover, the cosmological perturbations are far less problematic~\cite{DeFelice:2012mx}: all the d.o.f. propagate at the linear level without ghost instabilities~\cite{Comelli:2012db}. For more recent works see \cite{BergAkrami}.

Another problem which arises in massive gravity models is generic for theories with extra propagating degrees of freedom.
Since the graviton mass turns on  (at least) three extra degrees of freedom, it is expected that the extra interaction change the Newtonian limit or/and the light deflection. This can be easily seen in the so-called decoupling limit --- the scalar part of the graviton is directly coupled to the matter with approximately the same coupling constant of the helicity-2 piece. 
The extra scalar behaves similar to the Brans-Dicke field ruling out the theory on observational ground. 
Moreover, a naive way to recover GR sending the mass of the graviton to zero, does not solve the problem --- the so called 
vDVZ discontinuity~\cite{vanDam:1970vg} --- 
since the theory with arbitrary small (but non-zero) graviton mass contains the extra propagating scalar, absent in the massless theory (GR). 
A way to overcome this difficulty was proposed by Vainshtein~\cite{Vainshtein:1972sx} in 1972. 
Vainshtein noticed that the linear approximation breaks down at some distance far from the source 
(now called the Vainshtein radius) and therefore one cannot approximate the solution by linearizing it close to the source. 
On the contrary, he showed that a solution can be found by expanding around the GR solution in powers of the graviton mass. 
This construction indicates the possibility that the extra propagating scalar mode can be hidden close to the source by non-linear effects. The question still remained if this close-to-GR solution could have been matched to an asymptotically flat solution~\cite{Boulware:1973my};
it was in fact argued that it was not possible~\cite{Jun:1986hg,Damour:2002gp}.
Only recently it has been realized that the Vainshtein solution close to the source matches the one obtained by linearization far from the source.
Therefore GR is restored locally for asymptotically flat solutions, at least for some potentials 
in massive gravity~\cite{Babichev:2009us,Babichev:2010jd}\footnote{For the DGP model~\cite{Dvali:2000hr} the cosmological version of the Vainshtein mechanism was found in~\cite{Deffayet:2001uk}.}
(see also \cite{Alberte:2010it} for a more recent work).
This matching was shown for potentials giving rise to the sixth dangerous mode.
Later it was found, both analytically in the so-called decoupling limit~\cite{Koyama:2011xz,Koyama:2011yg,Chkareuli:2011te} and numerically~\cite{Volkov:2012wp}, 
that GR is also restored in the dRGT model. 
In the framework of the same model, the Vainshtein mechanism was studied in the decoupling limit for asymptotically non-flat spacetimes in~\cite{Berezhiani:2013dw}, where these solutions were shown to be the only stable ones
(see also a related work on Galileons~\cite{Babichev:2012re} and on Horndeski theory \cite{Koyama:2013paa}); and in the quasi-dilaton extension in~\cite{D'Amico:2012zv}.
For a recent review on the \V mechanism see~\cite{Babichev:2013usa}.\footnote{For completeness we mention that, very recently, a full class of new massive gravity potentials has been found
in the Lorenz Breaking scenario \cite{Comelli:weak}; these models do not suffer of the vDVZ discontinuity and therefore do not need to rely on the \V mechanism to recover GR.} 
Other possible issues of massive gravity we are not going to discuss here include superluminality~\cite{superluminality} 
(see, however, discussion in~\cite{Burrage:2011cr, Bruneton:2006gf} on relation between causality and superluminality),
strong coupling problem~\cite{ArkaniHamed:2002sp, Burrage:2012ja}, instability of black holes~\cite{Babichev:2013una}.

For what concerns the bi-gravity formulation of the dRGT model,
the numerical study of the Vainshtein mechanism was presented in~\cite{Volkov:2012wp}, while
the far-distance analytic expansion valid outside the Vainshtein radius was found in~\cite{Comelli:2011wq} and then studied up to the second order in~\cite{Enander:2013kza}.
Some estimates of the Vainshtein suppression have been put forward also in~\cite{DeFelice:2013nba} 
in order to calculate the emission of gravitations waves.
However, there is still a lack of an analytic analysis of the Vainshtein mechanism, that we fill in this work.
 
In this paper we analytically study the spherically symmetric solutions in the bi-gravity extension of the dRGT model
and show that the Vainshtein mechanism indeed works also with a dynamical second metric. 
We make our analysis in the approximation of the weak gravitational field of the physical metric --- the one coupled to the matter --- which is 
always correct for (nonrelativistic) weak matter sources. This assumption allows us to treat the problem mostly analytically. 
The key equation we obtain is an algebraic polynomial equation, generically of the 7th order, for a function of the radius entering, in our ansatz,  in the second metric. The other functions that parameterize the metrics are expressed in terms of this key function. 
In two regimes --- inside and outside the Vainshtein radius --- we can solve (approximately) the algebraic equation that 
gives different branches of the solution, then we identify the one which ensures the flat asymptotic behavior. 
We exhibit a solution featuring the GR behavior for the physical metric inside the Vainshtein radius 
and that matches the asymptotically flat (Yukawa decaying) solution of the linearized equations. 

The paper is organized as follows. 
In Sec.~\ref{dRGT model} for completeness,  we reanalyze the Vainshtein mechanism for the original dRGT model (with fixed reference metric)
in a slightly different manner than in~\cite{Koyama:2011xz,Koyama:2011yg,Chkareuli:2011te}. 
This approach will be generalized in the main part of the paper (Sec.~\ref{bi-gravity model}) for the bi-gravity extension, where we rigorously study the Vainshtein mechanism with the second metric dynamic.
Our conclusions are formulated in Sec.~\ref{Conclusion}.

\section{dRGT model}\label{dRGT model}
In this section we obtain static spherically symmetric solutions in the limit of weak gravitational field for the dRGT model.
We reproduce the results already found in~\cite{Koyama:2011xz,Koyama:2011yg,Chkareuli:2011te} 
for distances smaller than the Compton wavelength of the graviton, 
where the decoupling limit (DL) is a good approximation. 
In our approach however, originally introduced in~\cite{Babichev:2010jd} and called ``weak-field approximation'', 
the DL scheme is not used.
The weak-field approximation allows to capture both the DL and the Yukawa part of the solution outside the Compton wavelength, 
where the DL ceases to operate.\footnote{In particular, we derive an ODE for the gauge function which is valid at all radii, in the limit of weak source. 
For practical purposes this equation is not useful, however, it may show some important features, e.g. 
for the dRGT potential compared to a generic one see discussion in Sec.~\ref{SSSSm}.}
In Sec.~\ref{bi-gravity model}, this scheme --- with appropriate modifications --- will be applied to the bi-gravity extension of the model.

\subsection{Action and equations of motion}
The action for the dRGT model can be written as follows~\cite{deRham:2010ik}
\be
S =  M^2_P\int d^4 x \sqrt{-g} \left(\frac{R[g]}{2} + m^2 {\cal U} [g,f]\right) 	+ S_m [g].
\ee
It is convenient to express the interaction potential ${\cal U}[g,f]$ in terms of the matrix $\mathcal{K}$, such that $\mathcal{K}^\mu_\nu = \delta^\mu_\nu - \gamma^\mu_\nu$, 
where the matrix $\gamma^\mu_\nu$ is the square root of the product of the inverse physical metric $g^{\mu\alpha}$ and the fiducial metric $f_{\alpha\nu}$,
i.e. $\gamma^\mu_\nu = \sqrt{g^{\mu\alpha}f_{\alpha\nu}} $, in the sense that 
$\left(\gamma^2\right)^\mu_\nu = \gamma^\mu_\alpha\gamma^\alpha_\nu = g^{\mu\alpha}f_{\alpha\nu}$. As it is often assumed we will consider the fiducial 
metric to be flat\footnote{We use the mostly positive signature (-+++).}. The potential $\cal{U}$ consists of three pieces, 
\be\label{U234}
	\U = \U_2 + \alpha_3 \U_3 + \alpha_4 \U_4,
\ee
each of them, in terms of $\mathcal{K}$, reads
\begin{equation}
\label{UdRGT}
\begin{aligned}
	\U_2 &= \frac{1}{2!}\left( [\K]^2 - [\K^2] \right), \\
	\U_3 &= \frac{1}{3!}\left( [\K]^3 -3 [\K] [\K^2] +2[\K^3]\right), \\
	\U_4 &= \frac{1}{4!}\left( [\K]^4 -6 [\K]^2 [\K^2]+3 [\K^2]^2 + 8[\K^3][\K] -6[\K^4]\right)\,,
\end{aligned}
\end{equation}
where we introduced the notations $[\mathcal{K}]\equiv \text{tr}(\hat{K}) = \hat{\mathcal{K}}^\rho_\rho$ and 
	$[\mathcal{K}^n]\equiv \text{tr}(\hat{\mathcal{K}}^n) = (\hat{\K}^n)^\rho_\rho $.

Varying the action with respect to $g_{\mu\nu}$ one obtains
\begin{equation}
	G_{\mu\nu} = m^2 T_{\mu\nu} + \frac{T_{\mu\nu}^{(\text{m})}}{\M2},\nonumber
\end{equation}
where $G_{\mu\nu}$ is the Einstein tensor and on the right hand side there are the contributions from the energy-momentum tensor for the matter, 
$T^{(m)}_{\alpha\beta}\equiv -\frac{2}{\sqrt{-g}}\frac{\delta S}{\delta g^{\alpha\beta}}$,
and for the interaction term with $f_{\m\n}$, 
\begin{equation*}
	T_{\mu\nu} = \U g_{\mu\nu} - 2 \frac{\delta \U}{\delta g^{\mu\nu}} .
\end{equation*}
The last can be computed and gives 
\begin{equation}
\begin{aligned}
	T_{\mu\nu} =& - g_{\mu\sigma}\gamma^{\sigma}_{\alpha}\left(\K^\alpha_\nu -[\K]\delta^\alpha_\nu\right)
			+\alpha_3 g_{\mu\sigma}\gamma^{\sigma}_{\alpha}\left(\U_2\delta^\alpha_\nu - [\K]\K^\alpha_\nu+(\K^2)^\alpha_\nu\right)\\
			&+\alpha_4 g_{\mu\sigma}\gamma^{\sigma}_{\alpha}\left(\U_3\delta^\alpha_\nu-\U_2\K^\alpha_\nu + [\K](\K^2)^\alpha_\nu-(\K^3)^\alpha_\nu\right) +\U g_{\mu\nu}.\nonumber
\end{aligned}
\end{equation}

\subsection{Static Spherically Symmetric Solutions}\label{SSSSm}
In this section we study spherically symmetric solutions for the case where the non-dynamical second metric $f_{\m\n}$ parametrizes a flat Minkowski space-time. 
The study of the Vainshtein mechanism in this case has been already done in a number of papers~\cite{Koyama:2011yg, Chkareuli:2011te, Sbisa:2012zk}, 
in the decoupling limit.
Here we reproduce these results, moreover we will give some additional
new upshots outside the DL regime.
The procedure is to consider the full equations of motion and then make reasonable approximations valid for the regimes in which we are interested.
Following the weak-limit approximation scheme~\cite{Babichev:2010jd}, we take the following ansatz:
\be
\begin{split}
	ds^2 &= - e^{\nu}dt^2 + e^{\lambda}dr^2 + r^2d\Omega^2,\\
	df^2  &= - dt^2 + \left(r+ r\mu\right)'^2  dr^2 +  (r+r\mu)^2d\Omega^2.
\end{split}	\label{ansatz1}
\ee
This ansatz is not the most general, indeed we do not consider the case with one of the two metrics off-diagonal,
but~(\ref{ansatz1}) is where we find the \V mechanism at work.

Since we are interested in a recovery of GR solutions,
we require for weak matter sources to have weak gravity, i.e. that the functions $\nu$ and $\lambda$ are small as well as their derivatives.
So the first step is to consider:
\be
	\{\lambda,\nu\} \ll 1, \quad \{r\lambda',r\nu'\}\ll 1 \,, \label{wfr}
\ee
and to retain all the non-linearities in $\mu$ and $\mu'$.
The $tt$, $rr$ and $\theta\theta$ components of the Einstein equations in this approximation read,
\bea
	- \frac{\lambda'}{r} - \frac{\lambda}{r^2} &=& 
	m^2\left(\frac{\lambda}{2} +\frac{1}{r^2}\left\{r^3\left(-\mu+\alpha\mu^2-\frac{\beta}{3}\mu^3\right)\right\}'\right) - \frac{\rho}{\M2},\label{Einsteintt}\\
	\frac{\nu'}{r}-\frac{\lambda}{r^2} &=& m^2\left(\frac{\nu}{2} - 2\mu + \alpha\mu^2 \right)\label{Einsteinrr}\\
	- \frac{\lambda'}{2r} + \frac{1}{2}\left(\nu''+\frac{\nu'}{r}\right) &=& m^2\left(\frac{\nu+\lambda}{2} 
	+ \frac{1}{r}\left\{r^2\left(-\mu+\frac{\alpha}{2}\mu^2\right) \right\}' \right).\label{Einsteinthetatheta}
\eea
where we introduced
\be
	\alpha = 1+\alpha_3, \quad \beta = \alpha_3 + \alpha_4.\nonumber
\ee
The Bianchi identity, $\nabla_\mu T^\mu_r = 0$, gives 
\be
\label{Bianchi}
	-\frac{\lambda}{r}+\frac{\nu'}2+\alpha\left(\frac{\lambda}{r}-\nu'\right)\mu + \frac{\beta}2 \nu'\mu^2 = 0 \,.
\ee
Note that the pressure in the r.h.s. of (\ref{Einsteinrr}) and (\ref{Einsteinthetatheta}) disappears 
as a consequence of the conservation of the matter energy momentum tensor in the weak field regime (\ref{wfr}).
Of course, like in GR, the three Einstein equations and the Bianchi one are not all independent, so we can consider
(\ref{Einsteintt}), (\ref{Einsteinrr}) and (\ref{Bianchi}) as our independent set to be solved.

From this set we are able to obtain one second order ODE on~$\mu$ only.
Indeed we can solve (\ref{Einsteinrr}) for $\lambda$ and then, substituting into (\ref{Einsteintt}) and (\ref{Bianchi}),
we end up with two equations, one for $\nu', \nu, \mu$, and the other for $\nu'', \nu', \nu, \mu', \mu$. Taking the first equation and its first and second derivative, together with the second equation and its first derivative, we have a system of five equations in $\nu''', \nu'', \nu', \nu, \mu'', \mu', \mu$. We can then solve algebraically four of them for $\nu$ and all its three derivatives and, substituting in the last equation, it will be a second order differential equation only on $\mu$ of the form
\be
\label{wfe}
{\cal A}\,\m'' + {\cal B}\,{\m'}^3 + {\cal C}\,{\m'}^2 + {\cal D}\,\m' + {\cal E} = 0 \,.
\ee
$\cal A, B, C, D, E$ are functions of $\mu$ whose form is not particularly illuminating, so we can omit it here.
It is worth to stress the difference with respect to the equivalent equation found in the ghosty massive gravity theories~\cite{Babichev:2010jd}. 
Our second order equation needs two initial conditions in order to be solved, fixing to one the number of degree of freedom that it describes. 
For other kind of potentials~\cite{ArkaniHamed:2002sp} that exhibits the BD instability, the equation is of the fourth order~\cite{Babichev:2010jd}, 
meaning four initial conditions and therefore two d.o.f. One of these modes is absent for the dRGT potential.

Eq.~(\ref{wfe}) is hard to solve, even numerically. 
Without solving it, however, the equation clearly indicates that the weak field approximation~(\ref{wfr})
gives the relevant features of the fields for all ranges of distance for non-relativistic sources.
In order to understand the behavior of the solutions, below we will consider various regimes.

\subsubsection{Linear regime}

Since we require asymptotically flat solutions, we expect that far away from the source also the field $\mu$ becomes
small.
Therefore, assuming $\mu \ll 1$ in~(\ref{Einsteintt}),~(\ref{Einsteinrr}) and (\ref{Bianchi}), as well as its derivative, we get the solutions:
\bea
	\mu &=& - \, \frac{C\, e^{-m\,r}}{3\,m^2\,r^3} \left[1+m\,r \left(1 + m\,r \right) \right] \,, \label{mulin} \\
	\lambda &=& \frac{2\,C\, e^{-m\,r}}{3\,r} \left(1+m\,r\right) \,, \label{lambdalin}\\
	\nu &=& - \, \frac{4\,C\, e^{-m\,r}}{3\,r} \,, \label{nulin}
\eea
where $C$ is an integration constant that in the following we will see equal to the \sch radius $r_S$.
Clearly the gravitational potentials (\ref{lambdalin}) and (\ref{nulin}) exhibit the vDVZ discontinuity in the limit of $m \rightarrow 0$\,, 
instead the field $\mu$ shows a singularity in the same limit.
Actually the linear regime is nowhere allowed in the vanishing mass limit, indeed this regime exists only for values of radii for which $\mu \ll 1$. 
For non-zero small $m$, as can be easily seen from~(\ref{mulin}), the linear regime is valid for $r \gg r_V$,
where $r_V$ is the \V radius,
\be
r_V \equiv \left( \frac{r_S}{m^2} \right)^{1/3} \,.\nonumber
\ee
In the limit $m \rightarrow 0$ we have $r_V \rightarrow \infty$\,, making unreliable the condition to be outside $r_V$.
Hence, to look at the small mass limit, we need to consider a non-linear regime in $\mu$.
Finally, it is remarkable that the weak field approximation is able to retain the asymptotically Yukawa decay at large distances from the source, as well the \V crossover as we will see in the next paragraph.

\subsubsection{Inside the Compton wavelength}

In order to study the behavior of the solutions for which the $m \rightarrow 0$ limit is well defined and the \V mechanism operates, 
we need to consider the distances inside the Compton wavelength, i.e. $r\ll 1/m$.
In this regime we can neglect terms~$\sim m^2 \lambda$ and $\sim m^2 \nu$ in the r.h.s. of equations (\ref{Einsteintt})-(\ref{Einsteinthetatheta}). In this approximation, we can integrate equation (\ref{Einsteintt}) to obtain
\be
\label{EinsteinttV}
\lambda=
\begin{cases}
 \frac{r_S}{r} + m^2\,r^2\left(\mu- \alpha\,\mu^2 + \frac{\beta}{3}\,\mu^3\right) & \qquad \text{for}\,\,\, r > R_\odot  \\[.5cm]
 \frac{\rho\, r^2}{3\, \M2} + m^2\,r^2\left(\mu- \alpha\,\mu^2 + \frac{\beta}{3}\,\mu^3\right) & \qquad \text{for}\,\,\,  r < R_\odot
\end{cases}
\ee
where $R_\odot$ is the radius of the source, the Schwarzschild radius $r_S$ reads
\be
	r_S = \frac{1}{\M2}\int_0^{R_\odot} \rho\, r^2 dr \,,\nonumber
\ee
and  the integration constant in (\ref{EinsteinttV}) has been chosen to insure the continuity of the solution at the surface of the star.
In the following for simplicity we will consider only a constant density source.
Neglecting~$\sim m^2 \lambda$ and $\sim m^2 \nu$ in Eq.~(\ref{Einsteinrr}) one obtains
\be
\frac{\nu'}{r} -\frac{\lambda}{r^2} = m^2\left( - 2\mu + \alpha\mu^2 \right), \label{EinsteinrrVV}
\ee
while the integration of (\ref{Einsteinthetatheta}) gives (\ref{EinsteinrrVV}) up to an integration constant. 
From (\ref{EinsteinrrVV}) and (\ref{EinsteinttV}) we find
\be
\label{EinsteinrrV}
r\nu' =
\begin{cases}
 \frac{r_S}{r} - m^2\,r^2\left(\mu- \frac{\beta\,\mu^3}3\right) & \qquad \text{for}\,\,\, r > R_\odot  \\[.5cm]
 \frac{\rho\, r^2}{3\, \M2} - m^2\,r^2\left(\mu- \frac{\beta\,\mu^3}3\right) & \qquad \text{for}\,\,\,  r < R_\odot
\end{cases}
\ee
Finally, combining (\ref{Bianchi}), (\ref{EinsteinttV}) and (\ref{EinsteinrrV}) we get a single algebraic equation on $\mu$,
\be
\label{mueq}
3\,\mu  - 6\,\alpha\,\mu^2 + 2\left(\alpha^2+\frac{2\,\beta}{3}\right)\mu^3 - \frac{\beta^2}3\mu^5 = 
\begin{cases}
- \frac{r_S}{m^2\,r^3}\left(1-\beta\mu^2\right) & \qquad \text{for}\,\,\, r > R_\odot  \\[.5cm]
- \frac{\rho}{3\,m^2\,\M2}\left(1-\beta\mu^2\right) & \qquad \text{for}\,\,\,  r < R_\odot
\end{cases}
\ee
The last equation~(\ref{mueq}) corresponds to the one found in the DL of the model~\cite{Koyama:2011yg, Chkareuli:2011te, Sbisa:2012zk}. This confirms that the approximations we made here correspond to the DL in the full equations of motion.

All the physics is hence enclosed in equation (\ref{mueq}): once we have its solutions we can determine the gravitational potentials through equations (\ref{EinsteinttV}) and~(\ref{EinsteinrrV}).
Eq.~(\ref{mueq}) is a fifth order algebraic equation and its solutions can not be presented in a closed form.
Choosing the parameters of the theory $\a$ and $\b$, we can lower the degree of such equation in order to get analytically solvable ones.
E.g., for the minimal massive gravity potential with only $\U_2$\, --- 
it corresponds to $\b=0$ and $\a=1$ --- Eq.~(\ref{mueq}) becomes a third order algebraic equation. 
This special case was first studied in~\cite{Koyama:2011xz}, where it was shown that the \V mechanism works properly reproducing GR inside the \V radius.

Since the solutions of the set of equations (\ref{mueq}), (\ref{EinsteinttV}) and~(\ref{EinsteinrrV}) have been 
largely studied~\cite{Koyama:2011yg, Chkareuli:2011te, Sbisa:2012zk}, we only report schematically the behavior of the solutions for different sub-regimes.
To consider the most interesting case, for which the \V mechanism takes place, we set $\b>0$ so that
Eq.~(\ref{mueq}) has two complex and three real solutions.
Only one of the three real branches of the solution recovers GR and is asymptotically flat, 
so we give it in Table~(\ref{tablemono}).

\be\label{tablemono}
\begin{array}{c|c|c|c}
\hline
\quad r \quad &\quad r < R_\odot \quad & \quad R_\odot < r \ll r_V \quad & \quad r_V \ll r \ll 1/m \\
\hline
\quad \m \quad & \quad - \frac{1}{\sqrt \beta} + \frac{m^2\, R_\odot^3\left(\a^2 + 3\,\a\,\sqrt \b +2\, \b \right)}{\beta^2\, r_S} \quad & \quad - \frac{1}{\sqrt \beta} + \frac{m^2\, r^3\left(\a^2 + 3\,\a\,\sqrt \b +2\, \b \right)}{\beta^2\, r_S} \quad & - \frac{r_S}{3\,m^2\,r^3} \\
\quad \l \quad & \frac{r_S\,r^2}{R_\odot^3} - \frac{m^2\, r^2\left(3\,\a + 4\,\sqrt \b \right)}{3\, \beta} & \frac{r_S}{r} - \frac{m^2\, r^2\left(3\,\a + 4\,\sqrt \b \right)}{3\, \beta} & \frac{2\,r_S}{3\,r}  \\
\quad \n \quad &  - \frac{3\,r_S}{2\,R_\odot} + \frac{r_S\,r^2}{2\,R_\odot^3} + \frac{m^2\,r^2}{3\,\sqrt \b} & - \frac{r_S}{r} + \frac{m^2\,r^2}{3\,\sqrt \b} &  - \frac{4\,r_S}{3\,r} \\
\hline
\end{array}
\ee
Inside the \V radius we find that $\mu = - 1/\sqrt\b$ at the leading order and the gravitational potentials are of the GR form plus small corrections. 
Clearly, the vanishing mass limit is well defined inside the \V radius and the corrections to GR smoothly vanish as $m\to 0$.
Outside the \V radius we find the asymptotically flat weak field solution where also $\mu \ll 1$:
this solution match the one obtained in the linear regime (\ref{mulin})-(\ref{nulin}) provided $C=r_S$ and $r \ll 1/m$. 

To get some understanding of the solutions from Eq.~(\ref{mueq}), note the r.h.s of itself.
The ratio $(r_V/r)^3$ that appears there becomes large inside the \V radius, and small otherwise.
For $r \ll r_V$ we have either $|\mu| \gg 1$ 
retaining so the higher power of $\mu$ in the equation, 
or the leading order of $\mu$ cancels the r.h.s. itself, i.e.~$\mu = \pm 1/\sqrt\b$.
For $r \gg r_V$ instead we can neglect at the first order the r.h.s. of~(\ref{mueq}),
obtaining therefore three real constant values for~$\mu$: obviously only $\mu=0$ gives asymptotically the flat metric.

The branch of the solution presented in~(\ref{tablemono}) 
is unphysical for $\beta<0$ due to the square root in $\mu$. 
A complete description of the solutions for the whole range of the free parameters $\a$ and $\b$ can be found in~\cite{Sbisa:2012zk}.

\section{bi-gravity}\label{bi-gravity model}
This section is devoted to the study of the \V mechanism in the Hassan-Rosen bi-gravity extension of the dRGT model~\cite{Hassan:2011zd}. 
In the bi-gravity approach the second metric~$f_{\mu\nu}$, which was fixed before, now becomes dynamical.
The weak-field approximation, that we applied in the previous section, is also useful here. 
The trick is to deform the ansatz (\ref{ansatz1}) to include the dynamics of the metric $f_{\mu\nu}$.
After writing down the action and the field equations in Sec.~\ref{actionbi}, we introduce the ansatz and identify 
the functions that can be treated linearly in the limit of weak sources, together with a fully non-perturbative function.
In this way we are able to obtain again one algebraic polynomial equation,
which captures the distances inside the Compton wavelength of the graviton.
\subsection{Action and equations of motion}\label{actionbi}
We consider the dRGT model where 
the additional metric $f_{\mu\nu}$ is dynamical thanks to its own Einstein-Hilbert term in the action~\cite{Hassan:2011zd},
\begin{equation}
\label{actionBI}
	S = M^2_P\int d^4 x\sqrt{-g}\left( \frac{R[g]}{2} + m^2 \U[g,f] \right) +S_m[g] + \frac{\kappa M^2_P}{2}\int d^4 x \sqrt{-f} \R[f].
\end{equation}
The interaction potential in (\ref{actionBI}) is given by the same expressions in~(\ref{U234}) and (\ref{UdRGT})\footnote{Note that 
in principle we can add terms $\U_0 =1$ and $\U_1 = [\mathcal{K}]$. These terms, however, account for cosmological terms for the $g$ and $f$ metric. 
Since we aim to find asymptotically flat solutions, we exclude those terms.}.
Note an extra parameter $\kappa$ in the action~(\ref{actionBI}), which 
accounts for a possible difference in Planck masses for the two gravitational sectors.
One can realize that the limit $\kappa\to \infty$ corresponds to the freezing of the metric $f$, therefore recovering the model with flat fiducial metric.
Varying the action (\ref{actionBI}) with respect to $g_{\mu\nu}$, we obtain
\begin{equation}
	G_{\mu\nu} = m^2 T_{\mu\nu} + \frac{T_{\mu\nu}^{(\text{m})}}{\M2},\nonumber
\end{equation}
where $G_{\mu\nu}$ is the Einstein tensor associated with $g_{\mu\nu}$ and $T_{\mu\nu}$ reads,
\begin{equation}
\begin{aligned}
	T_{\mu\nu} &= \U g_{\mu\nu} - 2 \frac{\delta \U}{\delta g^{\mu\nu}} = \\
			& - g_{\mu\sigma}\gamma^{\sigma}_{\alpha}\left(\K^\alpha_\nu -[\K]\delta^\alpha_\nu\right)
			+\alpha_3 g_{\mu\sigma}\gamma^{\sigma}_{\alpha}\left(\U_2\delta^\alpha_\nu - [\K]\K^\alpha_\nu+(\K^2)^\alpha_\nu\right)\\
			&+\alpha_4 g_{\mu\sigma}\gamma^{\sigma}_{\alpha}\left(\U_3\delta^\alpha_\nu-\U_2\K^\alpha_\nu + [\K](\K^2)^\alpha_\nu-(\K^3)^\alpha_\nu\right) +\U g_{\mu\nu}.\nonumber
\end{aligned}
\end{equation}
On the other hand, the variation of the action with respect to $f_{\mu\nu}$ gives,
\begin{equation}
	\sqrt{-f}\,\mathcal{G}_{\mu\nu} =  \sqrt{-g}\frac{m^2}{\kappa} \mathcal{T}_{\mu\nu}, \nonumber
\end{equation}
where
\begin{equation}
\begin{aligned}
	\mathcal{T}_{\mu\nu} & = - 2 \frac{\delta \U}{\delta f^{\mu\nu}} =\\
			&  f_{\mu\sigma}\gamma^{\sigma}_{\alpha}\left(\K^\alpha_\nu -[\K]\delta^\alpha_\nu\right)
			-\alpha_3 f_{\mu\sigma}\gamma^{\sigma}_{\alpha}\left(\U_2\delta^\alpha_\nu - [\K]\K^\alpha_\nu+(\K^2)^\alpha_\nu\right)\\
			&-\alpha_4 f_{\mu\sigma}\gamma^{\sigma}_{\alpha}\left(\U_3\delta^\alpha_\nu-\U_2\K^\alpha_\nu + [\K](\K^2)^\alpha_\nu-(\K^3)^\alpha_\nu\right).\nonumber
\end{aligned}
\end{equation}
One can observe a useful relation when working with up-down indices, namely, for $T^\mu_\nu \equiv g^{\mu\alpha}T_{\alpha\nu}$ and 
$\mathcal{T}^\mu_\nu \equiv f^{\mu\alpha}\mathcal{T}_{\alpha\nu}$, we have
$
	\mathcal{T}^\mu_\nu = -T^\mu_\nu + \U\delta^\mu_\nu.
$

\subsection{Static Spherically Symmetric Solutions}
%
Continuing the idea of the weak-field approximation scheme, we consider the following parametrization for the two metrics
\be
\begin{split}
	ds^2 &= - e^{\nu}dt^2 + e^{\lambda}dr^2 + r^2d\Omega^2,\\
	df^2  &= - e^{n}dt^2 + e^{l}\left(r+ r\mu\right)'^2  dr^2 +  (r+r\mu)^2d\Omega^2 \,,
\end{split}	
\label{ansatz2}
\ee
where $\nu,\, \l,\, n,\, l$ and $\mu$ are the $r$-dependent functions that describe the spherically symmetric foliation of the space-time in a common coordinate system.
Again, this ansatz is not the most general one, since we are considering bi-diagonal metrics.

Let us note that our ansatz (\ref{ansatz2}) is compatible with the bi-flat solution
\be
g=f=\eta=\text{diag}(-1,1,1,1) \,,
\label{bkg}
\ee
obtained when one imposes $T=\mathcal{T}=0$, and that 
will be the reference point for the asymptotic behavior.

\subsubsection{Static spherically symmetric ansatz and equations of motion.}

Following the procedure used for the dRGT model, we consider a non-relativistic matter source with constant density.
This allow us to assume $\nu$, $\lambda$, $n$ and $l$ small, as well as their derivatives,
\be
	\{\lambda ,\, \nu ,\, l , \, n \} \ll 1, \quad \{r\,\lambda', \, r\,\nu' , \, r\, l' , \, r\,n' \}\ll 1\,, \label{biwfr}
\ee
and to retain all the non-linearities in the field $\mu$ and $\mu'$\,.
The $tt$, $rr$ and $\theta\theta$ components of the Einstein equations in this approximation are for
the first metric 
\bea
	- \frac{\lambda'}{r} - \frac{\lambda}{r^2} &=& 
	m^2\left[ \frac12\left( \lambda - l \right) +\frac{1}{r^2}\left\{r^3\left(-\mu+\alpha\mu^2-\frac{\beta}{3}\mu^3\right)\right\}'\right] - \frac{\rho}{\M2}\,,\label{biEinsteintt}\\
	\frac{\nu'}{r}-\frac{\lambda}{r^2} &=& m^2\left[\frac12\left( \nu - n \right) - 2\mu + \alpha\mu^2 \right] \,, \label{biEinsteinrr}\\
	- \frac{\lambda'}{2r} + \frac{1}{2}\left(\nu''+\frac{\nu'}{r}\right) &=& m^2\left[\frac12
	\left( \nu + \lambda - n - l \right) 
	+ \frac{1}{r}\left\{r^2\left(-\mu+\frac{\alpha}{2}\mu^2\right) \right\}' \right]\,; \label{biEinsteinthetatheta}
\eea
and for the second metric
\bea
	&&- \left( 1+ \mu \right)\frac{l'}{r} - \left( r + r\, \mu \right)'\frac{l}{r^2} = 
	\frac{m^2}{\kappa}\left[\frac12\left( l - \lambda \right) +\frac{1}{r^2}\left\{r^3\left(\mu+\left(1-\alpha\right)\mu^2+\frac{1-\a+\beta}{3}\mu^3\right)\right\}'\,\right]\,,\nonumber \\
	\label{biEinstein2tt}\\
	&&\left(1+ \mu\right)\,\frac{n'}{r} - \left( r + r\, \mu \right)'\frac{l}{r^2} = \frac{m^2}{\kappa} \left[\frac12\left( n -\nu\right) + 2\,\mu + \left(1-\alpha\right)\mu^2 \right]\left( r + r\, \mu \right)' \,,\label{biEinstein2rr}\\
	&&- \frac{l'}{2r} 
	+ \frac{1+ \mu}{2\left( r + r\, \mu \right)'}\, n''+
	\left[1+ \left(r + r\,\mu \right)\left\{ \left[ \left( r + r\, \mu \right)' \,\right]^{-1}\right\}'\,\right] \frac{n'}{2\,r} = \nonumber \\
	&& \qquad\qquad\qquad\qquad\qquad\qquad\qquad\qquad\quad
	 \frac{m^2}{\kappa}\left[\frac12\left( l + n - \lambda -\nu \right) 
	+ \frac{1}{r}\left\{r^2\left(\mu+\frac{1-\alpha}{2}\mu^2\right) \right\}' \,\right] \,.\label{biEinstein2thetatheta}
\eea
Of course, these equations are not all independent, indeed a combination of (\ref{biEinsteinthetatheta}) and (\ref{biEinstein2thetatheta}) can be obtained taking a suitable combination of (\ref{biEinsteinrr}) and (\ref{biEinstein2rr}) and its derivative.
The Bianchi identities, $\nabla_\mu^{(g)} T^\mu_r \propto \nabla_\mu^{(f)}\left( \sqrt{-g}\mathcal{T}^\mu_r/\sqrt{-f}\right) = 0$, give
\be
	\frac1r \left( r + r\, \mu \right)' \left( 1 - \alpha \, \mu \right) \left( \lambda - l \right)-\frac12
	\left( 1-2\,\alpha\,\mu + \beta\,\mu^2\right)\left[\left( r + r\, \mu \right)'\nu' - n'\right] =0 \, . \label{biBianchi}
\ee
Note that assuming $l=n=0$ in (\ref{biBianchi}) we get back~(\ref{Bianchi}), the Bianchi identity for the model with one dynamical metric.
It is worth to mention that we were not able to obtain an analogue of equation~(\ref{wfe}), because, applying the similar approach that we described there,
we find a system of linear equations which is not linearly independent.
Therefore, in order to find analytical solutions to this set of equations, we need to do other approximations. 
This means to look at more specific regimes inside the weak field one~(\ref{biwfr}).

\subsubsection{Linear regime}

Since asymptotically we want to find the bi-flat solutions (\ref{bkg}), we expect that far away from the source also the field $\mu$ becomes small.
Hence, assuming $\mu\ll 1$ and $r \mu' \ll 1$ in~(\ref{biEinsteintt}), (\ref{biEinsteinrr}), (\ref{biEinstein2tt}), (\ref{biEinstein2rr}) and (\ref{biBianchi}), we find the solutions:
\bea
	\mu &=& -\, \frac{C_2 \, \kappa\, e^{-\, m\,r\,\sqrt{1+\frac{1}{\kappa}}}
   \left[\kappa +m\, r \left(m\, r \, (1+\kappa)+ \sqrt{\kappa  \left(1+\kappa\right)}
   \right)\right]}{3\,m^4 \,r^3 \left(1+\kappa\right)} \,, \label{bimulin} \\
	\lambda &=& \frac{C_1}{r} + \frac{2\,C_2 \, \kappa\, e^{-\, m\,r\,\sqrt{1+\frac{1}{\kappa}}}
   \left[\kappa +m\, r \sqrt{\kappa  \left(1+\kappa\right)}
   \right]}{3\,m^2 \,r \left(1+\kappa\right)} \,, \label{bilambdalin}\\
	\nu &=& - \frac{C_1}{r} - \frac{4\, C_2 \, \kappa^2\, e^{-\, m\,r\,\sqrt{1+\frac{1}{\kappa}}}
   }{3\,m^2 \,r \left(1+\kappa\right)} \,, \label{binulin} \\
   l &=& \frac{C_1}{r} - \frac{2\,C_2 \, e^{-\, m\,r\,\sqrt{1+\frac{1}{\kappa}}}
   \left[\kappa +m\, r \sqrt{\kappa  \left(1+\kappa\right)}
   \right]}{3\,m^2 \,r \left(1+\kappa\right)} \,, \label{billin}\\
	n &=& - \frac{C_1}{r} + \frac{4\,C_2 \, \kappa\, e^{-\, m\,r\,\sqrt{1+\frac{1}{\kappa}}}
   }{3\,m^2 \,r \left(1+\kappa\right)} \,, \label{binlin}
\eea
where $C_1$ and $C_2$ are two integration constants that will be determined in the next paragraph to be
\be
C_1= \frac{r_S}{1+\k} \,, \qquad\qquad C_2 = \frac{m^2\, r_S}{\k} \,. \label{C12}
\ee
It is important to stress that (once (\ref{C12}) is taken into account) taking the limit $\k \rightarrow \infty$, which freezes the dynamics of the second metric, we recover the solutions found in the same regime of the original dRGT model, i.e.  (\ref{mulin})-(\ref{nulin}) and $l=n=0$\,.

For the $m \rightarrow 0$ limit, the vDVZ discontinuity appears with the divergence in~$\mu$, the same arguments as in the previous section apply here: 
the  linear regime is nowhere allowed in the vanishing mass limit. Again, to properly consider this limit, we need to rely on nonlinearities in the field $\mu$\,.

\subsubsection{Inside the Compton wavelength}
\label{bivain}

As before, in order to study the \V mechanism we consider distances inside the Compton wavelength, i.e.~$r\ll 1/m$; 
this helps to avoid complications associated with the change of behavior at $r\sim 1/m$.
Neglecting therefore $\sim m^2 \lambda$ and $\sim m^2 l$ in the r.h.s. of (\ref{biEinsteintt}) and (\ref{biEinstein2tt}), we can integrate both the equations to obtain,
\be
\label{biEinsteinttV}
\lambda =
\begin{cases}
 \frac{r_S}{r} + m^2\,r^2\left(\mu- \alpha\mu^2 + \frac{\beta}{3}\mu^3\right) & \qquad \text{for}\,\,\, r > R_\odot  \\[.5cm]
\frac{\rho\, r^2}{3\, \M2} + m^2\,r^2\left(\mu- \alpha\mu^2 + \frac{\beta}{3}\mu^3\right) & \qquad \text{for}\,\,\,  r < R_\odot
\end{cases}
\ee
and
\be
\label{biEinstein2ttV}
l = 
 - \frac{m^2\,r^2}{\kappa\, \left( 1+\mu\right)} \left[ \mu+\left(1-\a \right)\mu^2 +\frac13\left(1-\a + \beta\right)\mu^3 \right],
\ee
where the integration constants have been chosen requiring the continuity of the solutions  at the surface of the star.
Similarly, neglecting also $\sim m^2 \nu$ and $\sim m^2 n$\, in ~(\ref{biEinsteinrr}) and (\ref{biEinstein2rr}) one finds,
%
\be
\label{biEinsteinrrV}
r\,\nu' = 
\begin{cases}
\frac{r_S}{r} - m^2\,r^2\left(\mu- \frac{\beta\mu^3}3\right) & \qquad \text{for}\,\,\, r > R_\odot  \\[.5cm]
 \frac{\rho\, r^2}{3\, \M2} - m^2\,r^2\left(\mu- \frac{\beta\mu^3}3\right) & \qquad \text{for}\,\,\,  r < R_\odot
\end{cases}
\ee
and
\be
\label{biEinstein2rrV}
r\,n' =\frac{m^2\,r^2\left( r + r\, \mu \right)'}{\kappa\,\left( 1+ \mu \right)^2}
	\left( \mu+ 2\, \mu^2+ \frac{2-2\,\a -\beta}{3}\,\mu^3 \right)\,,
\ee
while the integration of (\ref{biEinsteinthetatheta}) and (\ref{biEinstein2thetatheta}) do not give new equations.
Finally, combining (\ref{biBianchi}), (\ref{biEinsteinttV}), (\ref{biEinstein2ttV}), (\ref{biEinsteinrrV}) and (\ref{biEinstein2rrV}) we obtain 
a single algebraic equation on $\mu$,
\bea
   &&3 (\kappa +1) \mu +6 \left(\kappa +1\right) \left(1-\a\right) \mu ^2 + \nonumber \\
   &&\frac13 \left[6 \left(\kappa +1\right)\a^2 -2 \left(18\, \kappa +17\right)\a + 4 \left(\kappa +1\right)\beta +9\, \kappa +10\right] \mu ^3 + \nonumber \\
   && \frac23 \left[6 \left(\kappa +1\right) \a^2 - \left(9\,\kappa +7\right)\a + 4 \left(\kappa +1\right) \beta +1 \right] \mu ^4
   + \nonumber \\
   && \frac13\left[2 \left(3\, \kappa +1\right)\a^2 - \left(\kappa +1\right)\beta^2 + 2\left(2\, \kappa +1\right)\beta -4\,\a\,\beta -2\,\a \right] \mu ^5 - \nonumber \\
   && \frac23\, \kappa\, \beta^2\, \mu^6 -\frac13\, \kappa\, \beta^2\, \mu^7 =
   \begin{cases}
- \frac{\kappa\, r_S}{m^2\,r^3}\left( 1+ \mu \right)^2\left(1- \beta\,\mu^2\right) & \qquad \text{for}\,\,\, r > R_\odot  \\[.5cm]
- \frac{\kappa\, \rho}{3\,m^2\,\M2}\left( 1+ \mu \right)^2\left(1- \beta\,\mu^2\right) & \qquad \text{for}\,\,\,  r < R_\odot
\end{cases}
\label{bimueq}
\eea
Notice that equations (\ref{biEinsteinttV}) and (\ref{biEinsteinrrV}) for $\lambda$ and $\nu'$ are the same found for the dRGT model, 
see Eqs.~(\ref{EinsteinttV}) and (\ref{EinsteinrrV}).
It is also important to stress that dividing Eq.~(\ref{bimueq}) by $\kappa\,\left(1+\mu\right)^2$ and taking the limit $\kappa \to \infty$ that freezes $f$, 
we recover the master equation (\ref{mueq}) of the previous section. 
The fields $l$ and~$n'$ given in (\ref{biEinstein2ttV}) and (\ref{biEinstein2rrV}) vanish in the same limit.

Again, all the information is retained in an algebraic equation for $\mu$ only: 
for bi-gravity this equation is of the seventh order compared to the fifth order equation for the dRGT model. 
Generically equation (\ref{bimueq}) has seven solutions that can be real or complex. 

For~$\beta > 0$, it has three real and two complex conjugates solutions for all values of $r$; 
the remaining two solutions are real inside some radius $r_*$ and join together in a complex conjugates pair for $r > r_*$, see~Fig.~\ref{figmu}. 
The value of $r_*$ depends on the choice of the free parameters $\alpha$ and~$\beta$. 
The three everywhere real solutions for $\mu$, shown in Fig.~\ref{figmu}, have different asymptotic behavior, 
however all three recover GR inside the \V radius. 
One of these three is asymptotically flat (Vainshtein-Yukawa solution), and the 
others (dashed) have non-flat asymptotics (solutions three and four in Appendix~\ref{biunphy}).
One of these last two solutions, (\ref{tablebi4}), may be of interest in the context of cosmology. 
The asymptotically non-flat solution and a possible match to a cosmological one, however, deserves a separate study and will be discussed elsewhere.

For $\beta < 0$  two of the three solutions which are real for $\beta>0$ become complex conjugates, therefore only one everywhere real solution is left. 
The other solutions show the same behavior as in the case $\beta>0$. 
For $\beta<0$ the only everywhere real solution is given by the solution three of Appendix~\ref{biunphy} and it does not show the asymptotically flat behavior nor the expected weak field solution inside the \V radius.
Therefore, in the following, we will consider only the case $\beta>0 $.
The special case $\beta =0$, that for $\a=1$ corresponds to the minimal massive gravity potential with only $\U_2$, will be presented in Appendix \ref{simpot}.

However we do not guarantee that solutions exist for the whole range of parameters space; 
i.e. we do not exclude that for some range of the parameters $\alpha$ and $\beta$
the asymptotic solution might not match the solution inside the \V radius.
This issue would require a complete analysis for each value of $\alpha$ and $\beta$ that is beyond the scope of this work.
Indeed, contrary to Eq.~(\ref{mueq}) of the dRGT model, where a symmetry allows to study easily 
the whole range of parameters space \cite{Sbisa:2012zk}, in Eq.~(\ref{bimueq}) we were not able to find a similar strategy that facilitates the scan of solutions for general $\a$ and~$\b$.

\begin{figure}
\centering
\subfigure[\, Solutions that, for $r/r_V=r_*/r_V\simeq 0.31$, join together in a complex conjugates pair.]{\includegraphics[width=8.9cm]{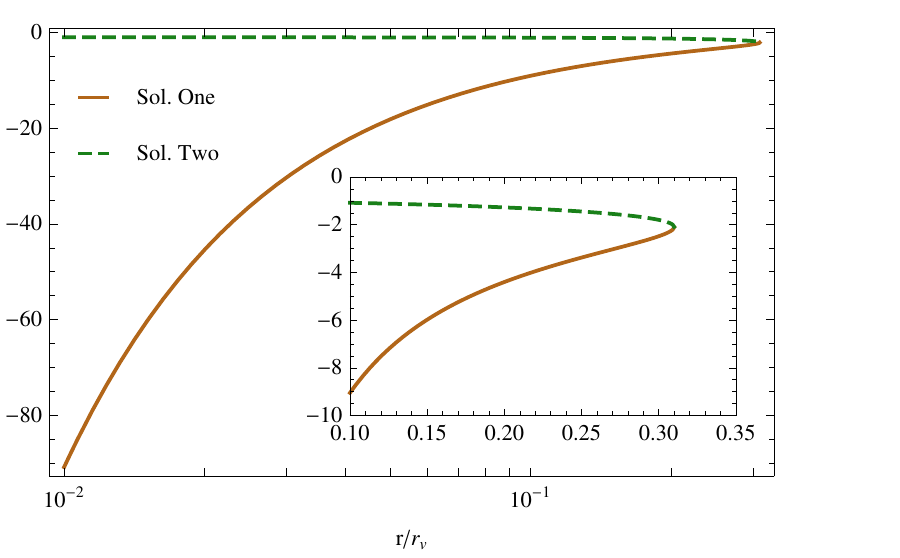}}
\subfigure[\, The three everywhere real solutions.]{\includegraphics[width=8.9cm]{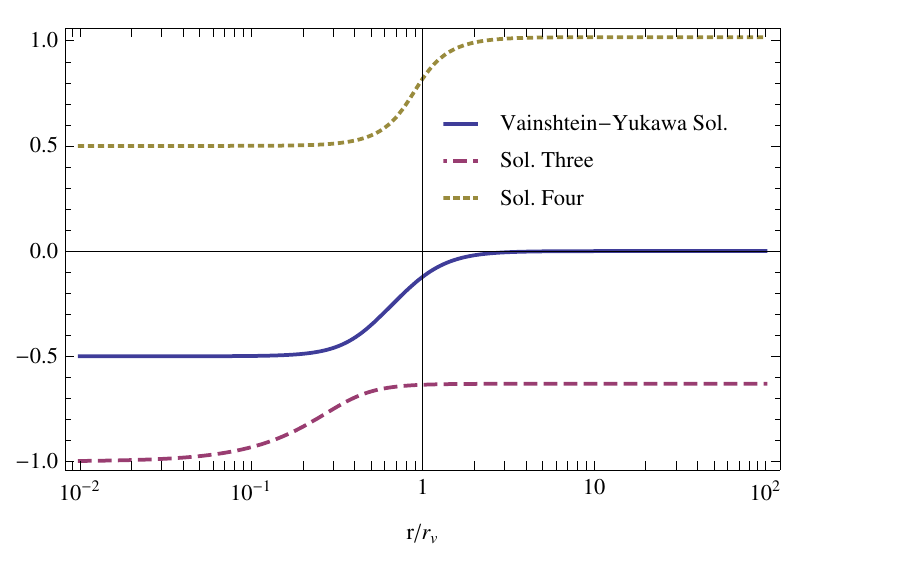}}
\caption{Plot of the five branches (out of  seven) of the solution of the function $\mu$ vs. $r/r_V$, for $R_\odot<r<1/m$. We take the following values for the parameters: $m \cdot r_V = 10^{-2},\, \k=1,\, \b=4$ and $\a=1$.
The name of each curve corresponds to the same name given in appendix~\ref{biunphy} where the analytic solutions are reported for the regime inside and outside the \V radious.
The Vainshtein-Yukawa solution is the one that reproduces GR inside $r_V$ and that gives the asymptotically flat solution outside $r_V$, as given in Tab~(\ref{tablebi}).}
\label{figmu}
\end{figure}

Again, one can understand the behavior of the solutions analyzing two regimes: well inside and outside the \V radius. 
For $r\ll r_V$ the ratio $(r_V/r)^3$ in the r.h.s. of (\ref{bimueq}) becomes large leaving us with two possibilities: 
either $|\mu| \gg 1$ in order to compensate the large r.h.s. (so at the leading order the higher powers of~$\mu$ dominate); 
or the leading order of $\mu$ cancels out the r.h.s., this happens for $\mu = - 1$ (double root) and~$\mu = \pm 1/\sqrt\b$.
For $r\gg r_V$, the ratio $(r_V/r)^3$ is small and it suggests to neglect at first order the r.h.s of (\ref{bimueq}): 
this determines the three real constant asymptotical values of $\mu$.  
One of these values is obviously zero and it is the one approached by the linear solution~(\ref{bimulin}).

In Table~(\ref{tablebi}) we present the branch of the solution which features the \V recovery of GR and the Yukawa decay.
The other branches are schematically presented in Appendix \ref{biunphy}.
The way how these regimes match together is understood from the numerical study that will be presented immediately after.

\be\label{tablebi}
\begin{array}{c|c|c|c}
\hline
\quad r \quad &\quad r < R_\odot \quad & \quad R_\odot < r \ll r_V \quad & \quad r_V \ll r \ll 1/m \\
\hline
\quad \m \quad & \qquad \hfill - \frac{1}{\sqrt \beta} + \d\mu \hfill \text{\underline{\tiny{$\d\mu \ll 1$}}} & \qquad \hfill - \frac{1}{\sqrt \beta} + \d\mu \hfill \text{\underline{\tiny{$\d\mu \ll 1$}}} & - \frac{r_S\, \kappa}{3\,m^2\,r^3 \left(1+ \kappa \right)} \\
\quad \l \quad & \frac{r_S\,r^2}{R_\odot^3} - \frac{m^2\, r^2\left(3\,\a + 4\,\sqrt \b \right)}{3\, \beta} & \frac{r_S}{r} - \frac{m^2\, r^2\left(3\,\a + 4\,\sqrt \b \right)}{3\, \beta} & \frac{r_S\left(3+2\,\k \right)}{3\,r\left(1+ \k \right)}  \\
\quad \n \quad &  - \frac{3\,r_S}{2\,R_\odot} + \frac{r_S\,r^2}{2\,R_\odot^3} + \frac{m^2\,r^2}{3\,\sqrt \b} &  - \frac{r_S}{r} + \frac{m^2\,r^2}{3\,\sqrt \b} & - \frac{r_S\left(3+4\,\k \right)}{3\,r\left(1+ \k \right)} \\
\quad l \quad & \quad - \frac{m^2\, r^2\left[1-\a - 3\left(1-\a\right)\sqrt \b + 4\, \b \right]}{3\, \beta\,\k \left(1 - \sqrt \b \right)} \quad & \quad - \frac{m^2\, r^2\left[1-\a - 3\left(1-\a\right)\sqrt \b + 4\, \b \right]}{3\, \beta\,\k \left(1 - \sqrt \b \right)} \quad & \frac{r_S}{3\,r\left(1+ \k \right)}  \\
\quad n \quad & \frac{m^2\, r^2\left[1-\a - 3\,\sqrt \b + \b \right]}{3\, \beta\,\k \left(1 - \sqrt \b \right)} & \frac{m^2\, r^2\left[1-\a - 3\,\sqrt \b + \b \right]}{3\, \beta\,\k \left(1 - \sqrt \b \right)} & \frac{r_S}{3\,r\left(1+ \k \right)} \\
\hline
\end{array}
\ee
\\
Inside the \V radius, for the metric $g$ the GR solution is recovered (plus small corrections) and 
for the second metric the potentials~$l$ and~$n$ are of the same order of the corrections to GR of the metric $g$. 
Outside the \V radius we find the asymptotically flat solution with $\mu \ll 1$:
this solution matches the one obtained in the linear regime~(\ref{bimulin})-(\ref{binlin}) for $r \ll 1/m$ and
\be
C_1= \frac{r_S}{1+\k} \,, \qquad\qquad C_2 = \frac{m^2\, r_S}{\k} \,, \nonumber
\ee
as we already anticipated in (\ref{C12}).
Note that inside the \V radius the vanishing mass limit, $m\to 0$, is well defined and gives exactly GR for the first metric while zero for $l$ and~$n$ potentials of the second metric.
It is also important to stress that in the limit $\kappa \to \infty$ for which $f_{\m\n}$ is frozen, this branch of the solution reproduces
the Vainshtein-Yukawa one of the dRGT model, namely Eq.~(\ref{tablemono}).

In addition to the Fig.~\ref{figmu}, in Fig.~\ref{figlambnu}, \ref{figln} and \ref{figint}
we depict correspondingly the potentials of the physical metric outside the source,
the potentials of the second metric outside the source and the potentials for both metrics inside the source,
only for the Vainshtein-Yukawa solution.

\begin{figure}[H]
\centering
\subfigure[\quad\quad]{\includegraphics[width=8.9cm]{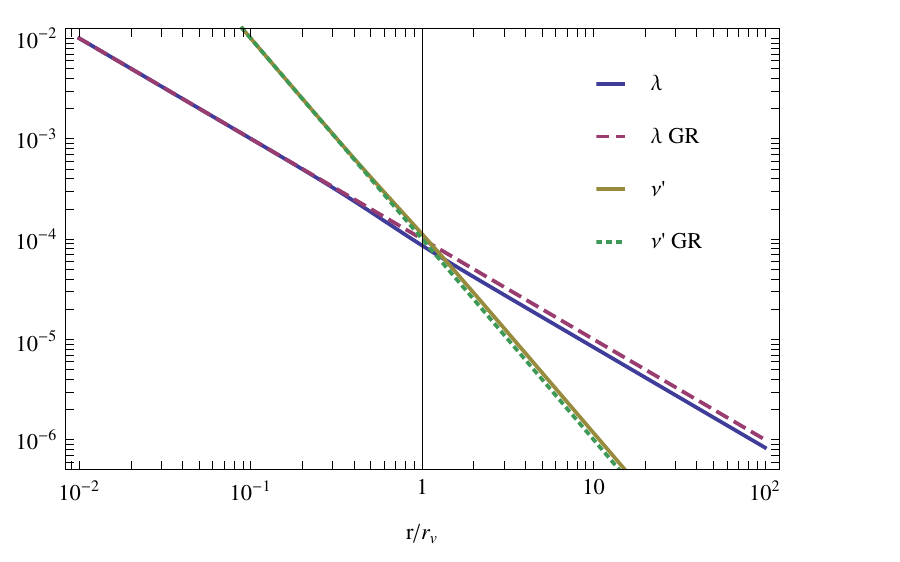}}
\subfigure[\quad\quad]{\includegraphics[width=8.9cm]{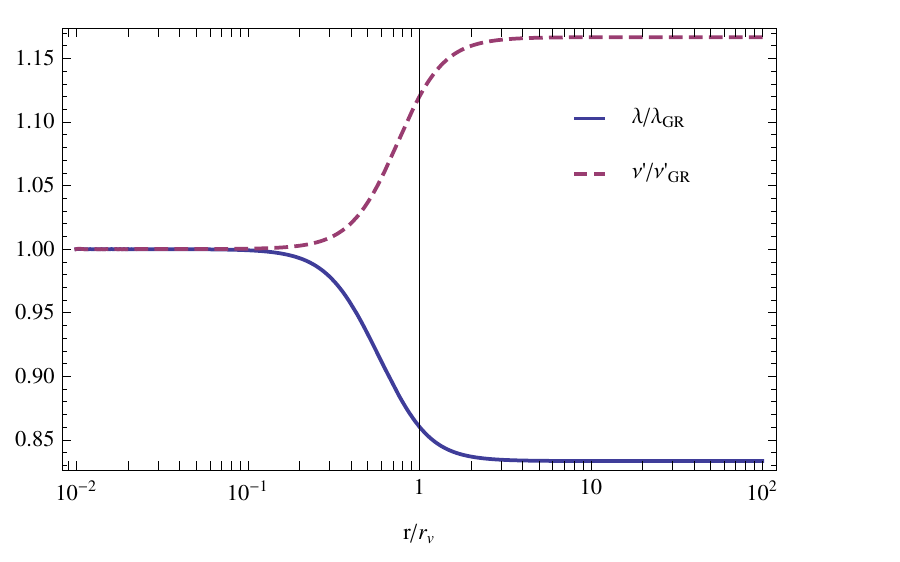}}
\caption{Plot of the Vainshtein-Yukawa solution of the functions $\l$ and $\n'$ vs. $r/r_V$, 
for $R_\odot<r<1/m$. We take the following values for the parameters: $m \cdot r_V = 10^{-2},\, \k=1,\, \b=4$ and $\a=1$.
These solutions are plotted together with the corresponding ones of GR in fig.~(a) and, in fig.~(b), is given their ratio for a better comparison.
The analytic behavior for the regime well inside and outside the \V radius is given in Tab.~(\ref{tablebi}).}
\label{figlambnu}
\end{figure}

\begin{figure}[H]
\centering
\includegraphics[width=10cm]{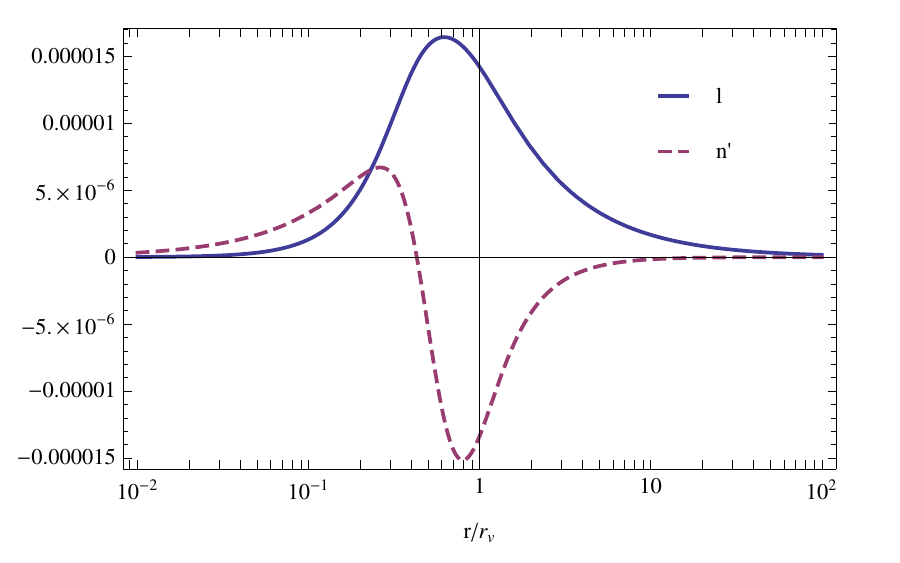}
\caption{Plot, for the Vainshtein-Yukawa branch, of the functions $l$ and $n'$ vs. $r/r_V$, for $R_\odot<r<1/m$. We take the following values for the parameters: $m \cdot r_V = 10^{-2},\, \k=1,\, \b=4$ and $\a=1$.
These numerical solutions show how the potentials $l$ and $n$ of the second metric get non-negligible values only in the intermediate regime around the \V radius. Indeed, as shown analytically in Tab.~(\ref{tablebi}), for $r\ll r_V$ their values are of the same order of magnitude of the corrections respect to GR of the potentials of the first metric and, for $r\gg r_V$, the solutions are asymptotically flat.}
\label{figln}
\end{figure}

\begin{figure}[H]
\centering
\includegraphics[width=10cm]{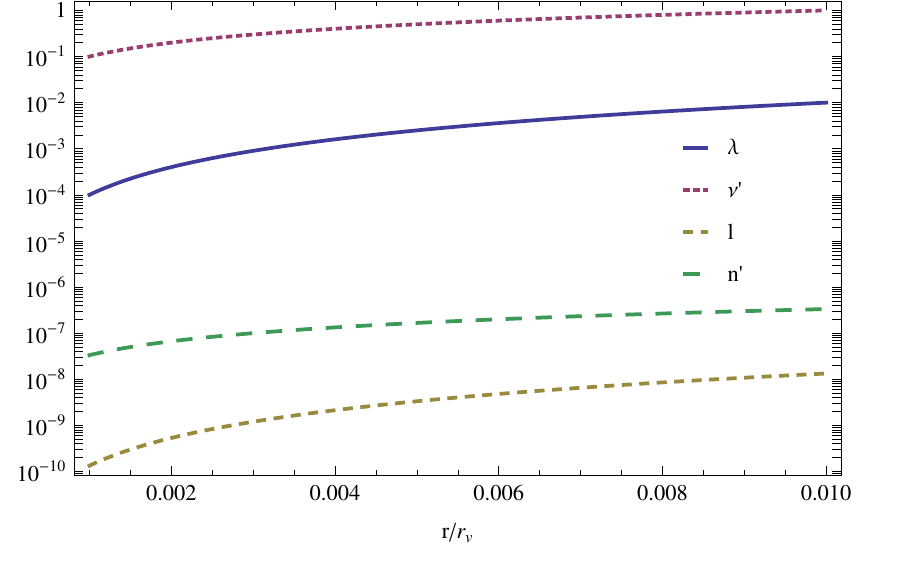}
\caption{Plot, for the Vainshtein-Yukawa branch, of the interior solution ($r<R_\odot$) for all the metrics functions (except $\m$ which is constant) vs. $r/r_V$. We take the following values for the parameters: $R_\odot=10^{-2}\,r_V,\, m \cdot r_V = 10^{-2},\, \k=1,\, \b=4$ and $\a=1$. The potentials of the first metric are indistinguishable from the analog in GR, instead the ones of the second metric are several orders of magnitude smaller. The analytic solutions are given in Tab.~(\ref{tablebi}).}
\label{figint}
\end{figure}

\section{conclusions}\label{Conclusion}
In this paper we studied the Vainshtein mechanism in the massive bi-gravity model with no Boulware-Deser ghost. 
To attack the problem, we applied the ``weak-field approximation scheme'' where 
the metric coefficients are separated into two parts. One contains functions of the radial coordinate which remain small 
(i.e. the quadratic and higher order terms are negligible in comparison to the linear ones) for non-relativistic sources, 
even inside the Vainshtein radius;
the other part is fully non-linear, with nonlinearities crucial for the existence of a GR like solution. 
This approach allows to capture all the important features of the solutions: the Vainshtein regime, the linear regime and the Yukawa decay.

In Sec.~\ref{dRGT model} we demonstrated how this scheme works for the original dRGT model --- 
where the auxiliary metric is fixed to Minkowski.
Inside the Compton wavelength, our results are in agreement with previous studies in the decoupling limit~\cite{Koyama:2011xz,Koyama:2011yg,Chkareuli:2011te}, see in particular~(\ref{tablemono}). 
On the other hand, we can also describe the solution outside the Compton wavelength, which is beyond the validity of the decoupling limit.
Moreover, for the function $\mu(r)$ which is introduced as the non-linear piece of the metric~(\ref{ansatz1}), 
we derived a single ordinary differential equation of the second order~(\ref{wfe}) valid for all radii. 

For bi-metric massive gravity we modified the approach to incorporate the dynamics of the second metric.
Notably, the ansatz we introduced~(\ref{ansatz2}), again separates the metric coefficients in the linearizable part
and in the fully non-linear part $\mu(r)$.
If we additionally assume that far from the source also $\mu$ is in the linear regime, then we readily obtain the linearized solution that shows the Yukawa decay.
Other regimes can be obtained assuming radii smaller than the Compton wavelength, i.e. $r\ll 1/m$. 
In this case, it is possible to derive one algebraic equation of the seventh order on $\mu$, Eq.~(\ref{bimueq}),
while the other metric functions are given in terms of it.
Using this master equation, we analyzed the behavior of the solutions in various sub-regimes and identified several branches of the solution. 
For $\beta>0$ the only solution that has the desired behavior ---  
asymptotic flatness --- is presented in the Table~(\ref{tablebi}): this solution shows the recovery of GR inside the Vainshtein radius.

It is worth to make a comment about the asymptotically non-flat solution (\ref{tablebi4}). 
Although we did not study (\ref{tablebi4}) in detail in the present work, since we concentrated on asymptotically flat solutions, 
this solution may have a physical meaning if matched to a cosmological solution at $r\to \infty$.
The same comment applies to the choice of the potential 
$\alpha_3 = \alpha_4 =0$,  which we discuss in Appendix~\ref{simpot}. 
While for the dRGT model this simplest potential gives an asymptotically flat solution recovering GR inside the Vainshtein radius,
in the case of the bi-metric massive gravity, such an asymptotic solution does not exist 
and the solution featuring the Vainshtein behavior becomes asymptotically non-flat.
The behavior and the physical meaning of these solutions, together with a possible match to cosmological ones, deserves a separate study.

To summarize, in the bi-gravity formulation of the dRGT massive gravity, with matter coupled to only one (physical) metric,
we have found the recovery of GR for the physical metric inside the Vainshtein radius and the Yukawa decay outside.
At the same time, the second metric is nontrivial because of the indirect coupling to matter via the interaction (mass) term; 
its deviation from flat spacetime is highly suppressed and it reaches non-negligible values only around the Vainshtein radius. 

\begin{acknowledgments}
We would like to thank Luigi Pilo and Michael Volkov for many interesting discussions and correspondence. 
The work of EB was supported in part by grant FQXi-MGA-1209 from the Foundational Questions Institute.
\end{acknowledgments}

\begin{appendix}

\section{Simplest massive gravity potential}
\label{simpot}

Here we present the study of the simplest massive gravity potential with only $\U_2$ in~(\ref{U234}). 
It corresponds to set $\b =0$ and $\a = 1$ in the equations of section \ref{bivain}. This case deserves a particular analysis both since 
it was the first one studied in the framework of the dRGT model~\cite{Koyama:2011xz}, showing a well working \V mechanism, 
and because the branch that realizes the Vainshtein-Yukawa solution (\ref{tablebi}) is not present for $\b=0$.
For these values of the free parameters, the master equation (\ref{bimueq}) becomes a fifth degree equation with only one real solution for all the range of distances; the other four solutions start (as $r$ increases) as two pairs of complex conjugates solutions and then divide into four real distinct ones, see Fig.~\ref{figmuspecial}. Let us thus concentrate on the everywhere real branch. 
Similar to the general case, we can analytically find solutions in different regimes, see Table~\ref{tablebispecial}.

\begin{savenotes} 
\be\label{tablebispecial}
\begin{array}{c|c|c|c}
\hline
\quad r \quad &\quad r < R_\odot \quad & \quad R_\odot < r \ll r_V \quad & \quad r_V \ll r \ll 1/m \\
\hline
\quad \m \quad & - \left(\frac{r_S}{2\,m^2}\right)^{1/3} \frac1R_\odot & - \left(\frac{r_S}{2\,m^2}\right)^{1/3} \frac1r & c_0 \\
\quad \l \quad & \frac{r_S\,r^2}{R_\odot^3} - \left(\frac{m\,r_S}{2}\right)^{2/3} \left(\frac{r}{R_\odot}\right)^2  & \frac{r_S}{r} - \left(\frac{m\,r_S}{2}\right)^{2/3} & \quad  \frac{r_S}{r} + m^2\,r^2\left(c_0 - c_0^2\right)  \\
\quad \n \quad & \quad c_\n\symbolfootnote[2]{$c_\n$ is an integration constant fixed by the matching condition to be: $c_\n = - \frac{3\,r_S}{2\,R_\odot} + \left(\frac{m^4\,r_S}{2}\right)^{1/3} \frac{R_\odot}{2}$.} + \frac{r_S\,r^2}{2\,R_\odot^3} + \left(\frac{m^4\,r_S}{2}\right)^{1/3} \frac{r^2}{2\,R_\odot} \quad & \quad - \frac{r_S}{r} + \left(\frac{m^4\,r_S}{2}\right)^{1/3} r \quad & - \frac{r_S}{r} - \frac12\,m^2\,r^2\,c_0 \\
\quad l \quad & -\frac{m^2\,r^2}{\k} & -\frac{m^2\,r^2}{\k} & - \frac{m^2\,r^2\,c_0}{\kappa\, \left( 1+c_0\right)}  \\
\quad n \quad & \frac{m^2\,r^2}{\k} & \frac{m^2\,r^2}{\k} & \frac{m^2\,r^2\,c_0 \,\left( 1+2\,c_0\right)}{2\,\kappa\, \left( 1+c_0\right)} \\
\hline
\end{array}
\ee
\end{savenotes}
Inside the \V radius we find that $|\m| \gg 1$ and, in contrast to the general case,  it is this behavior of $\mu$ that gives a well working \V mechanism with GR like solutions for the metric $g$ and small potentials $l$ and $n$ for the other metric. 
For the dRGT model with the simplest potential a similar recovery of GR was found in~\cite{Koyama:2011xz}, with flat asymptotic.
For the bi-gravity model we find that, for $r \gg r_V$,  $\mu$ asymptotically approaches a non-zero constant value $c_0$, 
giving non-decaying tails for all the other gravitational potentials. 
This solution does not match the asymptotically flat linear solution (\ref{bimulin})-(\ref{binlin}).
It does not mean, however, that the solution~(\ref{tablebispecial}) is non-physical, it may in fact match a nontrivial cosmological solution for large $r$.
This possibility, however, deserves a separate study and goes beyond the scope of this paper.
In Fig. \ref{figmuspecial} we show also the numerical study of this solution for all the range of distances inside the Compton wavelength.

\begin{figure}
\centering
\subfigure[\,  Absolute value of the all five solutions of the function~$\mu$~vs.~$r/r_V$.]{\includegraphics[width=8cm]{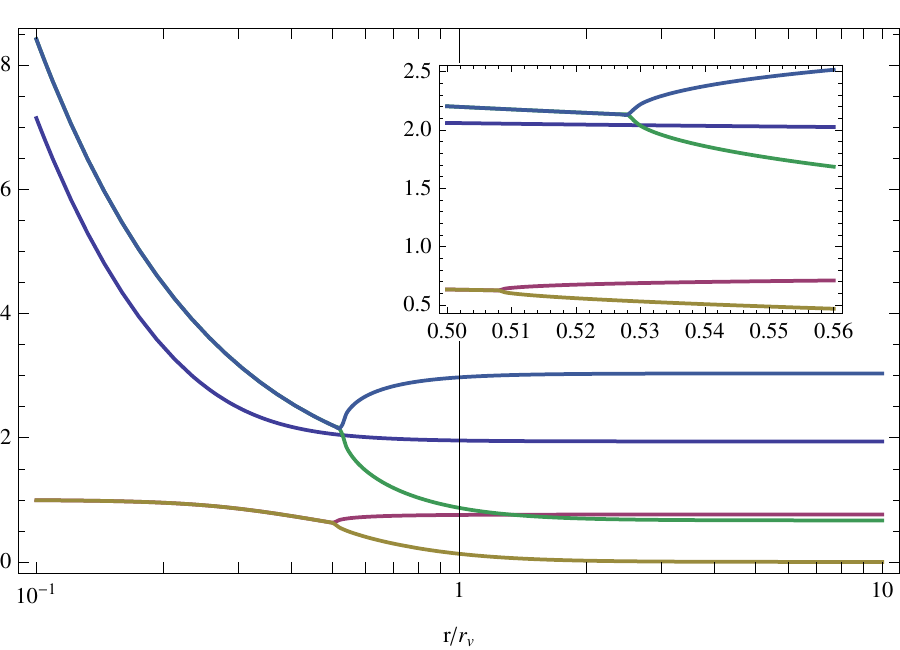}}
\subfigure[\, Functions $\m,\, \l$ and $l$ vs. $r/r_V$]{\includegraphics[width=9.5cm]{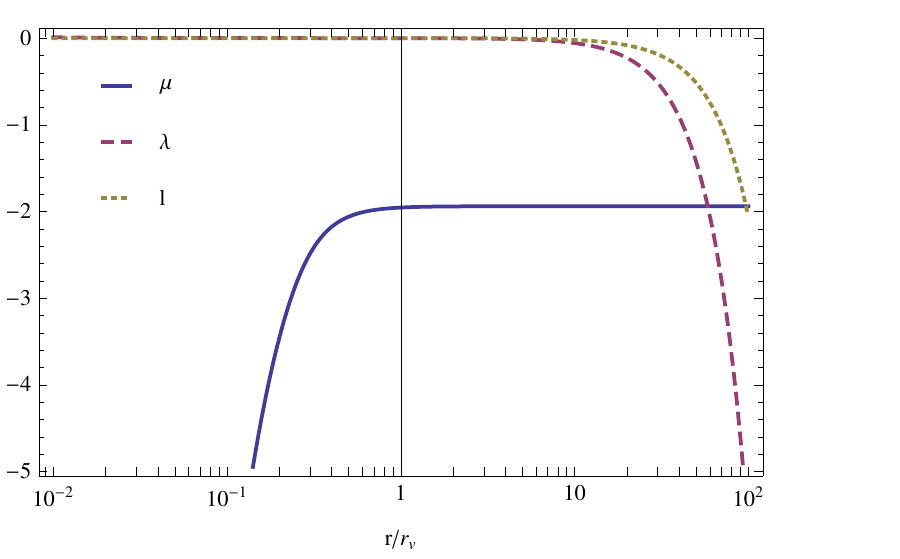}}
\caption{In fig. (a) plot of the absolute value of the all five solutions of the function $\mu$ vs. $r/r_V$, for $R_\odot<r<1/m$. There is only one everywhere real solution.
In fig. (b) plot of the functions $\m,\, \l$ and $l$ vs. $r/r_V$, for $R_\odot<r<1/m$.
As shown analytically in Tab~(\ref{tablebispecial}), the asymptotic behavior is not flat.
We take for both the plots the following values for the parameters: $m \cdot r_V = 10^{-2}$ and $\k=1$.}
\label{figmuspecial}
\end{figure}

The fact that the choice of parameters $\beta =0$ does not allow for an asymptotically flat everywhere real solution, 
while with the non-dynamical second metric such a solution exists, might seem surprising. 
There is however a simple explanation for this effect: when $\kappa \to \infty$ the outer part of the (nonphysical) asymptotically flat 
branch (dashed curve in Fig.~\ref{figmuspecialk}) and the inner part of the everywhere real solution (thick curve in Fig.~\ref{figmuspecialk}) 
join together to make one asymptotically flat solution recovering GR for small radii.
This can be easily seen from Fig.~\ref{figmuspecialk}, where the absolute value of the solutions of $\mu$ are plotted for bigger values of $\k$.

\begin{figure}
\centering
\subfigure[\,  $\k=10^2$]{\includegraphics[width=8cm]{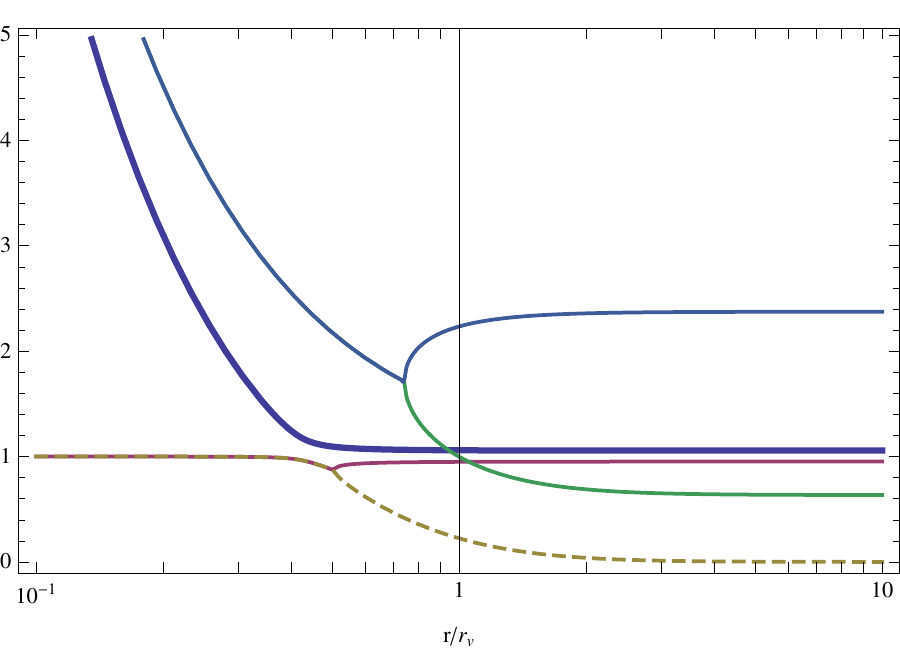}}
\subfigure[\, $\k=10^5$]{\includegraphics[width=8cm]{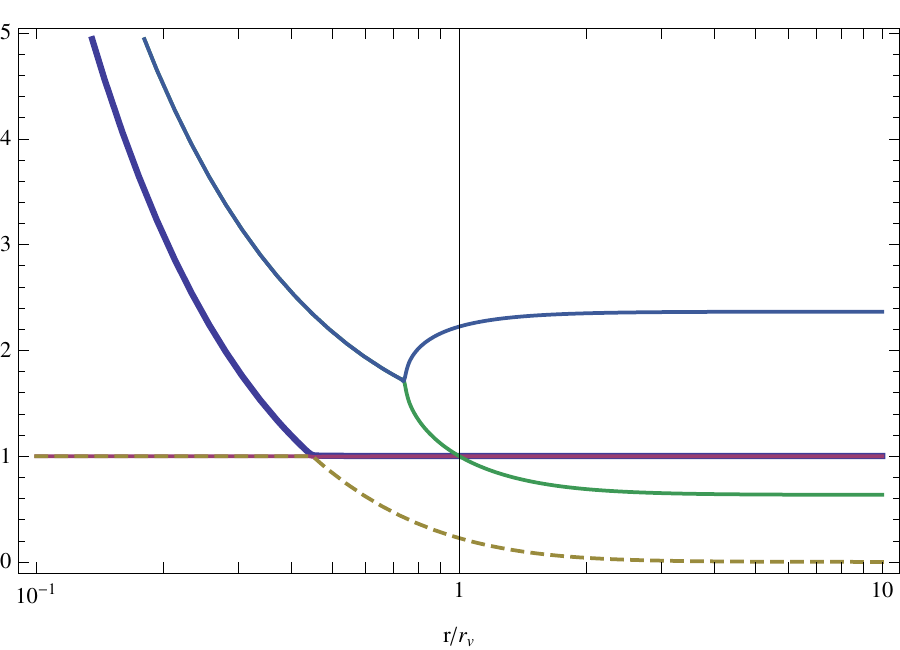}}
\caption{Plot of the absolute value of the all five solutions of the function $\mu$ vs. $r/r_V$, for $R_\odot<r<1/m$.
We take for fig. (a) $\k=10^2$ and for fig. (b) $\k=10^5$; both the plots have $m \cdot r_V = 10^{-2}$.}
\label{figmuspecialk}
\end{figure}

It is also instructive to compare our findings with the numerical work of Volkov~\cite{Volkov:2012wp}, 
where he obtains an asymptotically flat solution for the simplest potential ($\alpha=1$, $\beta =0$).
It looks as our results contradict the ones in~\cite{Volkov:2012wp}. 
In fact, a possible explanation lies in some specific choice of $m$, $r_S$ and $R_\odot$: 
it seems that in~\cite{Volkov:2012wp} these parameters are chosen such that 
the size of the source $R_\odot$ is larger than $r_*$ --- 
the point below which the asymptotically flat solution (dashed curve in~Fig.~\ref{figmuspecialk}) becomes complex --- 
therefore avoiding the problem since inside the source $\mu$ is constant. 
If the source is made more compact though still non-relativistic, we expect that this problem of complex-valued solution comes back,
although without complete numerical analysis of the full equations of motion we cannot prove this statement.

\section{Other  branches in massive bigravity}
\label{biunphy}

In this appendix we report the analytical solutions, for the regime inside and outside the \V radius, of the four branches of the massive bigravity solution that we omitted in the main text.
\\

{\bf Solution one}
\begin{savenotes}
\be\label{tablebi1}
\begin{array}{c|c|c|c}
\hline
\quad r \quad &\quad r < R_\odot \quad & \quad R_\odot < r \ll r_V \quad & \quad r_V \ll r \ll 1/m \\
\hline
\quad \m \quad & - \left(\frac{3\, r_S}{\beta\, m^2}\right)^{1/3} \frac1R_\odot  & - \left(\frac{3\, r_S}{\beta\, m^2}\right)^{1/3} \frac1r & // \\
\quad \l \quad & {\cal O} \left(\frac{1}{\mu}\right)^2 & {\cal O} \left(\frac{1}{\mu}\right)^2 & //  \\
\quad \n \quad &  {\cal O} \left(\frac{1}{\mu}\right)^2 &  {\cal O} \left(\frac{1}{\mu}\right)^2 & // \\
\quad l \quad & - \frac{1-\a+\b}{3^{1/3} \, \k} \left( \frac{m\,r_S}{\b} \right)^{2/3} \left(\frac{r}{R_\odot}\right)^2 & - \frac{1-\a+\b}{3^{1/3} \, \k} \left( \frac{m\,r_S}{\b} \right)^{2/3} & //  \\
\quad n \quad & \quad c_n\symbolfootnote[3]{$c_n$ is an integration constant fixed by the matching condition to be: $c_n = - \frac{2-2\,\a-\b}{2 \cdot 3^{2/3} \, \k} \left[ 3^{1/3} \left( \frac{m\, r_S}{\b} \right)^{2/3} + 2\, R_\odot \left( \frac{m^4\, r_S}{\b} \right)^{1/3} \right]$.} + \frac{2-2\,\a-\b}{2 \cdot 3^{1/3} \, \k} \left( \frac{m\, r_S}{\b} \right)^{2/3} \left(\frac{r}{R_\odot}\right)^2 \quad & \quad - \frac{2-2\,\a-\b}{3^{2/3} \, \k} \left( \frac{m^4\, r_S}{\b} \right)^{1/3}\, r \quad & // \\
\hline
\end{array}
\ee
\end{savenotes}
\\

{\bf Solution two}
\be\label{tablebi2}
\begin{array}{c|c|c|c}
\hline
\quad r \quad &\quad r < R_\odot \quad & \quad R_\odot < r \ll r_V \quad & \quad r_V \ll r \ll 1/m \\
\hline
\quad \m \quad & \qquad \hfill -1 + \d\mu \hfill \text{\underline{\tiny{$\d\mu \ll 1$}}} & \qquad \hfill -1 + \d\mu \hfill \text{\underline{\tiny{$\d\mu \ll 1$}}} &  // \\
\quad \l \quad & \frac{r_S\,r^2}{R_\odot^3} - m^2\, r^2 \left(1+\a + \frac{\b}{3} \right) & \quad \frac{r_S}{r} - m^2\, r^2 \left(1+\a + \frac{\b}{3} \right) \quad & //  \\
\quad \n \quad &  \quad - \frac{3\,r_S}{2\,R_\odot} + \frac{r_S\,r^2}{2\,R_\odot^3} + \frac12\, m^2\,r^2 \left( 1 - \frac{\b}{3} \right) \quad &  - \frac{r_S}{r} + \frac12\, m^2\,r^2 \left( 1 - \frac{\b}{3} \right)& // \\
\quad l \quad & {\cal O} \left(\frac{1}{\d\mu}\right) & {\cal O} \left(\frac{1}{\d\mu}\right) & //  \\
\quad n \quad & {\cal O} \left(\frac{1}{\d\mu}\right) & {\cal O} \left(\frac{1}{\d\mu}\right) & // \\
\hline
\end{array}
\ee
\\

{\bf Solution three}
\be\label{tablebi3}
\begin{array}{c|c|c|c}
\hline
\quad r \quad &\quad r < R_\odot \quad & \quad R_\odot < r \ll r_V \quad & \quad r_V \ll r \ll 1/m \\
\hline
\quad \m \quad & \qquad \hfill -1 + \d\mu \hfill \text{\underline{\tiny{$\d\mu \ll 1$}}} & \qquad \hfill -1 + \d\mu \hfill \text{\underline{\tiny{$\d\mu \ll 1$}}} & c_1 \\
\quad \l \quad & \frac{r_S\,r^2}{R_\odot^3} - m^2\, r^2 \left(1+\a + \frac{\b}{3} \right) & \quad \frac{r_S}{r} - m^2\, r^2 \left(1+\a + \frac{\b}{3} \right) \quad & \frac{r_S}{r} + m^2\,r^2\left(c_1- \alpha\, c_1^2 + \frac{\beta}{3}\,c_1^3\right)  \\
\quad \n \quad &  \quad - \frac{3\,r_S}{2\,R_\odot} + \frac{r_S\,r^2}{2\,R_\odot^3} + \frac12\, m^2\,r^2 \left( 1 - \frac{\b}{3} \right) \quad &  - \frac{r_S}{r} + \frac12\, m^2\,r^2 \left( 1 - \frac{\b}{3} \right) & - \frac{r_S}{r} - m^2\,r^2\,\frac12\left(c_1- \frac{\beta}{3}\,c_1^3\right) \\
\quad l \quad & {\cal O} \left(\frac{1}{\d\mu}\right) & {\cal O} \left(\frac{1}{\d\mu}\right) & \quad - \frac{m^2\,r^2}{\kappa\, \left( 1+c_1\right)} \left[ c_1+\left(1-\a \right)c_1^2 +\frac13\left(1-\a + \beta\right)c_1^3 \right]  \\
\quad n \quad & {\cal O} \left(\frac{1}{\d\mu}\right) & {\cal O} \left(\frac{1}{\d\mu}\right) & \frac{m^2\,r^2}{2\, \kappa\,\left( 1+ c_1 \right)}\left( c_1+ 2\, c_1^2+ \frac{2-2\,\a -\beta}{3}\,c_1^3 \right) \\
\hline
\end{array}
\ee
\\

{\bf Solution four}
\be\label{tablebi4}
\begin{array}{c|c|c|c}
\hline
\quad r \quad &\quad r < R_\odot \quad & \quad R_\odot < r \ll r_V \quad & \quad r_V \ll r \ll 1/m \\
\hline
\quad \m \quad & \qquad \hfill \frac{1}{\sqrt \beta} + \d\mu \hfill \text{\underline{\tiny{$\d\mu \ll 1$}}} & \qquad \hfill \frac{1}{\sqrt \beta} + \d\mu \hfill \text{\underline{\tiny{$\d\mu \ll 1$}}} & c_2 \\
\quad \l \quad & \frac{r_S\,r^2}{R_\odot^3} - \frac{m^2\, r^2\left(3\,\a - 4\,\sqrt \b \right)}{3\, \beta} & \frac{r_S}{r} - \frac{m^2\, r^2\left(3\,\a - 4\,\sqrt \b \right)}{3\, \beta} & \frac{r_S}{r} + m^2\,r^2\left(c_2- \alpha\, c_2^2 + \frac{\beta}{3}\,c_2^3\right)  \\
\quad \n \quad &  - \frac{3\,r_S}{2\,R_\odot} + \frac{r_S\,r^2}{2\,R_\odot^3} - \frac{m^2\,r^2}{3\,\sqrt \b} & - \frac{r_S}{r} - \frac{m^2\,r^2}{3\,\sqrt \b} & - \frac{r_S}{r} - m^2\,r^2\,\frac12\left(c_2- \frac{\beta}{3}\,c_2^3\right) \\
\quad l \quad & \quad - \frac{m^2\, r^2\left[1-\a + 3\left(1-\a\right)\sqrt \b + 4\, \b \right]}{3\, \beta\,\k \left(1 + \sqrt \b \right)} \quad & \quad - \frac{m^2\, r^2\left[1-\a + 3\left(1-\a\right)\sqrt \b + 4\, \b \right]}{3\, \beta\,\k \left(1 + \sqrt \b \right)} \quad & \quad - \frac{m^2\,r^2}{\kappa\, \left( 1+c_2\right)} \left[ c_2+\left(1-\a \right)c_2^2 +\frac13\left(1-\a + \beta\right)c_2^3 \right]  \\
\quad n \quad & \frac{m^2\, r^2\left[1-\a + 3\,\sqrt \b + \b \right]}{3\, \beta\,\k \left(1 + \sqrt \b \right)} & \frac{m^2\, r^2\left[1-\a + 3\,\sqrt \b + \b \right]}{3\, \beta\,\k \left(1 + \sqrt \b \right)} & \frac{m^2\,r^2}{2\, \kappa\,\left( 1+ c_2 \right)}\left( c_2+ 2\, c_2^2+ \frac{2-2\,\a -\beta}{3}\,c_2^3 \right) \\
\hline
\end{array}
\ee
\\

Let us give some comments on these branches.
The first two are given by the solutions of $\mu$ that at some point become complex conjugates (therefore the marks ``$//$'' for $r_V \ll r \ll 1/m$), so we can estimate their behavior only inside the \V radius. 
For the solution one,~$|\mu| \gg 1$ and this produces the screening of the gravitational potentials~$\l$ and $\n$ 
already seen in the dRGT model with one dynamical metric~\cite{Koyama:2011yg}.
For the solutions two and three, at the leading order $\mu=-1$ giving GR like solutions for the first metric potentials and very large values 
for the potentials of the second metric, i.e. $l,\,n \gg1$.
This is due to the fact that, for the ansatz~(\ref{ansatz2}), the inverse of the second metric is singular for $\mu$ strictly equal to $-1$.
It should be stressed that these solutions violate the assumption of weak field approximation (\ref{biwfr}), therefore they are not viable.
In branch three and four, outside the \V radius, $\mu$ approaches  
asymptotically to non-zero constant values $c_1$ and $c_2$:  this gives non-decaying gravitational potentials.
Finally, in the branch four, for $r \ll r_V$ we find that the \V mechanism works properly, recovering GR as in the Vainshtein-Yukawa branch~(\ref{tablebi}).

\end{appendix}



\end{document}